\documentclass[11pt, preprint]{article}
\usepackage{color} \usepackage{cancel}
\usepackage{float} \usepackage{graphicx}% Include figure files
\usepackage{graphicx}% Include figure files
\graphicspath{{images/}{images2/}}  % папки с картинками
\usepackage{subfigure} \usepackage{caption}	% Advanced figure caption
\usepackage{dcolumn}% Align table columns on decimal point
\usepackage[toc,page]{appendix} \usepackage{authblk}	 		% Title page
\usepackage{indentfirst}	% Indent for first paragraph
\usepackage{authblk}	 		% Title page
\usepackage{multirow}     % Multi-row cell in table
\usepackage{hyperref}     % Internal and external hyper links
\usepackage{amsfonts, amssymb, amsmath}
\usepackage[left=2.5cm,right=2.5cm,top=2.9cm,bottom=2.9cm,a4paper]{geometry}
\usepackage[noadjust]{cite}
\usepackage{filecontents}
\usepackage{soul}
\usepackage{xcolor}
\setstcolor{red}

\newcommand{\be}{\begin{equation}}
\newcommand{\ee}{\end{equation}}
\newcommand{\bea}{\begin{eqnarray}}
\newcommand{\eea}{\end{eqnarray}}

\graphicspath{{./figures/}}
\hypersetup{ colorlinks   = true, % Color links instead of ugly boxes
urlcolor     = blue, % Color of external hyperlinks
linkcolor    = blue, % Color of internal links
citecolor    = red 	 % Color of citations }
}

\title{\Large\bf A New Approach to the Analysis of Cosmological Parameters in Multifield Cosmology }
\author[1]{K. A. Bolshakova\thanks{bolshakova.ktrn@gmail.com}}
\author[1,2,3]{S. V. Chervon\thanks{chervon.sergey@gmail.com}}
\affil[1]{\small \it  Ulyanovsk State Pedagogical University, Ulyanovsk, Russia }
\affil[2]{\small \it Bauman Moscow State Technical University, Russia}
\affil[3]{\small \it Kazan Federal University, Kazan, Russia}

\begin{document}

\maketitle
%----------------------------------------------------------------
\begin{abstract}
Currently, a method has been developed for the cosmological inflation model with a single scalar field to calculate cosmological parameters such as the power spectrum of scalar and tensor perturbations, their spectral indices, and the tensor-to-scalar ratio. However, for a multifield configuration, a definitive method for calculating cosmological parameters has not yet been established. 

We propose a new effective algorithm for determining cosmological parameters within the tensor-multi-scalar theory of gravity, which describes the early inflationary epoch of the Universe. Our approach is based on utilizing a specific analytical solution within a multifield cosmological model to establish functional relationships between fields. This method enables the computation of cosmological parameters and their comparison with observational data.

\end{abstract}
%\tableofcontents
%----------------------Introduction------------------------------

\section{Introduction}

The theory of cosmological inflation emerged as a necessity to resolve critical issues in the Big Bang (BB) theory, such as the horizon, flatness, monopole, and galaxy formation problems. The inclusion of a scalar field in describing cosmological inflation during the early stages of the Universe's evolution facilitated significant progress in addressing these issues and provided mechanisms for the formation of the large-scale structure of the Universe based on quantum fluctuations during the inflationary epoch \cite{1s,2g,3L,4}. 

The system of equations for a self-gravitating model of a self-interacting scalar field is a complex nonlinear system of ordinary differential equations within the framework of a homogeneous and isotropic Universe. Consequently, exact solutions for scalar field models in cosmology were discovered nearly a decade later \cite{5,6,7,8}. Notably, based on exact solutions of scalar cosmology, cosmological perturbation parameters can be directly derived without analyzing linear perturbation equations \cite{9,10}.

Algorithms for calculating cosmological parameters necessary for alignment with observational data from WMAP and Planck satellites are well-developed within the framework of single scalar field models \cite{10,11}. Several methods exist for determining cosmological parameters such as the power spectra of scalar and tensor perturbations, their spectral indices, and the tensor-to-scalar ratio \cite{10, 12}. However, in multifield models, a specific method must be developed for each configuration. Successful generalizations of both exact solution construction methods (e.g., the superpotential method) and algorithms for calculating cosmological parameters for multifield models have been presented \cite{11,12,13}.

In general, two approaches can be distinguished for calculating cosmological perturbation parameters in multifield models: (1) using an ansatz for transitioning from single scalar fields to chiral fields \cite{12, 14}, and (2) employing linear relationships between fields \cite{13}.

The aim of this work is to propose a new approach based on utilizing a specific analytical solution in a multifield chiral cosmological model to establish functional relationships between fields. As the foundational theory of gravity, we consider a version of modified gravity, specifically the tensor-multi-scalar theory of gravity (TMS TG), introduced by Damour and Esposito-Farèse in 1992 \cite{16}.

In modern cosmology, modifications of Einstein’s gravity are extensively used to study the Universe's evolution, largely due to experimental evidence confirming the accelerated expansion of the Universe at present. This fact suggests that Einstein's theory of gravity cannot naturally explain the observed acceleration without introducing additional fields or exotic matter. Hence, there is a need to explore new modified gravity theories that, under certain approximations, include general relativity (GR). One such theory is TMS TG, which extends scalar-tensor gravity and represents a special case of chiral cosmological models \cite{17,18,19}. It is worth noting that \cite{19} examines a TMS TG model with multiple scalar fields interacting with gravity, for which solutions for the dust epoch, radiation dominance, and matter dominance have been derived. Investigations of cosmological inflation in TMS TG have been presented in \cite{20,21}, where solutions for three-field models in two inflationary scenarios—power-law and de Sitter—were obtained. This work builds upon the cosmological solutions derived in \cite{20}. 

\section{Cosmological Dynamics in TMS TG}

The action of the tensor-multi-scalar theory of gravity (TMS TG) with scalar fields included in the gravitational sector is given by \cite{16}:
\begin{equation} \label{EF}
		S= \frac{1}{\varkappa}\int d^4 x \sqrt{-g}\left[ \frac{R}{2}-\frac{1}{2}g^{\mu\nu}h_{AB} \varphi^{A}_{,\mu} \varphi^{B}_{,\nu}-W(\varphi^{C})\right] + S_m[\chi_m , \Omega^2(\varphi^{C})g_{\mu\nu}],
\end{equation}
Here, $\varkappa = 1$ is Einstein's gravitational constant, $R$ is the scalar curvature, and $g = \det (g_{\mu\nu})$. For brevity, we use $\varphi_{,\mu} = \partial_{\mu} \varphi$. Greek indices $\mu, \nu, \ldots = 0, 1, 2, 3$ denote spacetime coordinates, while uppercase Latin indices $A, B, C, \ldots = 1, 2, \ldots, N$ refer to $N$ scalar fields. The set of scalar fields $\left\lbrace \varphi^1, \varphi^2, \ldots, \varphi^N \right\rbrace$ will henceforth be denoted as $\varphi := \left\lbrace \varphi^1, \varphi^2, \ldots, \varphi^N \right\rbrace$, with $\chi$ being a scalar field.

The components of the internal metric $h_{AB}$ are presented as a two-component target space metric cosmological model  (CCM) with the metric:
	\begin{equation}\label{kir}
		d s_\sigma^2 = h_{11}d\phi^2 + h_{22}(\phi,\psi)d\psi^2,~ h_{11}=const.
	\end{equation}

The spacetime metric for a homogeneous and isotropic universe is taken as the Friedmann-Robertson-Walker (FRW) metric:
\begin{equation}\label{FRU}
		d s^{2}=-d t^{2}+a^{2}(t) \left( \frac{d r^{2}}{1-\epsilon r^{2}}+r^{2}d\theta^{2}+r^2 \sin^{2}\theta d\varphi^2 \right),
	\end{equation}
where $a(t)$ is the scale factor and $\epsilon$ determines the spatial curvature ($k = 0, \pm 1$). Instead of considering an open or closed Universe, we can remain within a spatially flat Universe filled with a scalar field and a perfect fluid with an equation of state $ p_{cur} = -3\rho_{cur}, \rho_{cur} = -\epsilon/(3a^2)$ \cite{23}.

The gravitational sector of action \eqref{EF} in the absence of matter $S_m$ corresponds to the CCM  in natural units where $\kappa=c=1$. Solutions obtained for CCM in prior works \cite{16, 17, 18} can thus be treated as vacuum solutions for TMS TG.

Considering the choice of the model with the chiral space \eqref{kir} and the action for the self-interacting scalar field, action \eqref{EF} can be written in the following form:
	\begin{equation}\label{1}
		S=  \int d^4 x \sqrt{-g} \left[ \dfrac{R}{2}-\frac{1}{2}g^{\mu\nu} ( h_{11} \phi_{,\mu}\phi_{,\nu} +h_{22}(\psi,\phi)) \psi_{,\mu}\psi_{,\nu}  -W(\psi,\phi)+  \left( - \dfrac{1}{2}g^{\mu\nu}  \chi_{,\mu} \chi_{,\nu} -U(\chi) \right) \right]. 
   	\end{equation}

As evident from action \eqref{1}, the model includes the following fields: chiral fields $\phi$ and $\psi$, and a single scalar field $\chi$ as the source of gravity. By varying action \eqref{1} with respect to the metric and the fields, we obtain the system of equations in the class of metrics (\ref{kir}), (\ref{FRU}) of the following form \cite{20,21}:

\begin{equation}\label{0b}
		3H\dot{\psi}h_{22}+\partial_t(h_{22}\dot{\psi})-\frac{1}{2}\frac{\partial h_{22}}{\partial \psi}\dot{\psi}^2+\frac{\partial W(\phi,\psi)}{\partial\psi}=\frac{\partial \ln \Omega(\phi,\psi)}{\partial
			\psi}(\dot\chi^{2}+4U(\chi)),
	\end{equation}
	\begin{equation}\label{1b}
		\ddot{\phi}h_{11}+3H\dot{\phi}h_{11} -\frac{1}{2}\frac{\partial h_{22}}{\partial\phi}\dot{\psi}^2 +\frac{\partial W(\phi,\psi)}{\partial\phi}=\frac{\partial \ln \Omega(\phi,\psi)}{\partial \phi}(\dot{\chi}^{2}+4U(\chi)),
	\end{equation}
	\begin{equation}\label{2b}
		H^2=\frac{1}{3}\left[ \frac{1}{2}h_{11} \dot{\phi}^2+\frac{1}{2}h_{22} \dot{\psi}^2+W(\phi,\psi)\right] +\frac{1}{3}\left(\frac{1}{2}\dot{\chi}^{2}+U(\chi)\right)-\frac{\epsilon}{a^2},
	\end{equation}
	\begin{equation}\label{3b}
		\dot{H}=-\left [\frac{1}{2}h_{11} \dot{\phi}^2+\frac{1}{2}h_{22} \dot{\psi}^2\right]-\dot{\chi}^{2} +\frac{\epsilon}{a^2},
	\end{equation}
	\begin{equation}\label{4b}
		\ddot{\chi}+3 H \dot{\chi}+U(\chi)_{,\chi}=0.
	\end{equation}
Here, the dot over a function represents a derivative with respect to cosmic time, $H=\frac{\dot{a}}{a}$ is the Hubble parameter, and $\Omega$ is the conformal transformation factor for transitioning from the Jordan frame to the Einstein frame: $g^E_{\mu \nu} = \Omega^2(x) g^J_{\mu \nu} $.
 	
The system of equations (\ref{0b}) -- (\ref{4b}) constitutes a fundamentally nonlinear system of equations describing the cosmological dynamics of the model under consideration (\ref{4b}). Following the approach proposed in \cite{16}, TMS TG is considered in the Einstein frame (without the nonminimal coupling of the scalar curvature to the scalar gravitational fields), where the matter field action as the source of gravity is considered in the "physical" metric $g^J_{\mu \nu}$, conformally related to the metric in the Einstein frame $  g^E_{\mu \nu} = \Omega^2(x) g^J_{\mu \nu} $. It is worth noting that under scaling $\Omega(\phi,\psi)=const$, the scalar field potential $U(\chi)$  influences the dynamics of the chiral fields $\phi$ and $\psi$ only through the Hubble parameter $H$.

The consequences of equations (\ref{2b}), (\ref{3b}) can be divided into equations for determining the kinetic and potential energy:

	\begin{equation}\label{5b}
		K(t)=\frac{1}{2}h_{11} \dot{\phi}^2+\frac{1}{2}h_{22}(\phi,\psi) \dot{\psi}^2+ \dot{\chi}^{2}= \frac{\epsilon}{a^2}-\dot{H},
	\end{equation}
	\begin{equation}\label{6b}
		W(t)=\left [\dot{H}+3H^{2}+ 2\frac{\epsilon}{a^2} -  U(\chi) \right].
	\end{equation}

    These equations were utilized in constructing decompositions (ansatzes) in \cite{6}. 
    
    In model \eqref{1}, there is additional freedom in choosing the conformal factor $\Omega(\phi,\psi)$. Following \cite{21}, we select $\Omega(\phi,\psi)$ in the following form:
\begin{equation}\label{om}
\Omega(\phi,\psi) = \exp (A\phi + B\psi), 
\end{equation}
where $A,B - const$.
Solutions for model \eqref{1} are presented in \cite{20,21}, where cosmological solutions are found for the case where the scalar field $\chi$ with a Higgs potential $U(\chi)$ is considered in the slow-roll regime. Power-law and exponential power-law evolution solutions for the scale factor were obtained for various limiting forms of the Higgs potential.

To make a transition to a single-field model, we select solutions corresponding to the case of a power-law scale factor $a(t) \propto t^m$ ($m > 1$) and a power-law potential $U(\chi) = \chi^k$, $D,k - const.$ \cite{20}.

\section{Method for Transitioning to a Single-Field Model}

The idea of the method involves expressing the dependence of the chiral fields $\phi$ and $\psi$ on the scalar field $\chi$ based on analytical solutions of the nonlinear system of equations (\ref{0b}) -- (\ref{4b}), using the inverse dependence $t(\chi)$. Then, knowing the dependencies $\phi(\chi)$ and $\psi(\chi)$, the potentials of the fields $ W (\psi,\phi)$ and  are expressed through the scalar field $\chi$. Additionally, the components of the chiral metric $h_{22}(\psi,\phi)$ are represented as functions of the scalar field $\chi (t)$: $h_{22} = h_{22}(\chi)$.

By substituting these functions into action \eqref{1}, which depends on $\chi$, we arrive at a single-field model.

Let us consider a specific analytical solution of the system (\ref{0b}) -- (\ref{4b}, where the matter field is treated under the slow-roll approximation, formally assuming $\dot{\chi}^2 \approx 0, \ddot{\chi} \approx 0 $.

\subsection{Transformation of TMS TG with Three Fields to a Single-Field Model}

For the power-law evolution of the scale factor $a(t) = t^m$, where $m= const$, $c=const$, and the self-interaction potential $U(\chi)=D\chi^k$, the solution obtained in \cite{20} is given by equation:
\begin{equation}\label{12}
		\chi=\left( \frac { Dk(k-2)}{6 m} t^2\right)^{\frac{1}{2-k}}, k \neq  2.
	\end{equation}

    In \cite{20}, the field $\chi$ corresponds to the field $\psi$. Solutions for chiral fields:
    	\begin{equation}\label{3}
		\psi=\sqrt{2}t,
	\end{equation}
	\begin{equation}\label{11}
		\phi=\sqrt{2m} \ln t.
	\end{equation}

    Let us express the chiral fields $\psi$ \eqref{3} and $\phi$ \eqref{11} in terms of the scalar field  $\chi$ \eqref{12}. To do this, we will determine the dependence of \( t (\chi) \) under the condition $k \neq 2$:
	\begin{equation} \label{t}
		t= \chi^{\frac{2-k}{2}} Q,
	\end{equation}
	where	$Q=\sqrt{ \frac { 6 m}{Dk (k-2)}}$. Next, substituting \eqref{t} into \eqref{3} and \eqref{11}, we obtain:
\begin{equation}\label{1.1}
		\phi(\chi) =\sqrt{2m} \ln\left ( Q\chi^{\frac{2-k}{2}}   \right),
	\end{equation}
	\begin{equation}\label{1.2}
		\psi(\chi) = \sqrt{2} Q\chi^{\frac{2-k}{2}}.  
	\end{equation}

To transition to a single-field model, it is also necessary to represent the components of the chiral metric in terms of the field \( \chi \). 
In our model, the first component of the chiral metric, \( h_{11}=1 \), does not depend on time. The second component of the chiral metric, \( h_{22} (\psi) \), is given by:
$ h_{22}(\psi) = \frac{\epsilon 2^m}{c^2 \psi^{2m}}$ .
Considering the relation \eqref{1.2}, we obtain:
	\begin{equation}\label{1.3}
		h_{22}(\chi)=\frac{ \epsilon \chi^{m(k-2)}}{c^2 Q^{2m}},  
	\end{equation}
where \( c= const\) is a constant corresponding to the power-law evolution of the scale factor \( a(t) = c t^m \).

To determine the total potential of all fields \( V_G(\chi) \), we utilize the potentials of the chiral fields found in \cite{20} for \( k \neq 2 \), taking into account the conformal factor \eqref{om}:

\begin{equation}\label{13}
		W_1(\phi)= m(3m-1) \exp \left({-\phi \sqrt{\frac{2}{m}}}\right),
	\end{equation}
	\begin{equation}\label{14}
		W_2 (\phi)= 4 \kappa D A \left( \frac { Dk(k-2)}{6 m}\right)^{\frac{k}{2-k}} \left( \frac{2-k}{2k}\right) \exp \left[ \frac{2k\phi}{\sqrt{2m } (2-k)} \right],
	\end{equation}
	\begin{equation}\label{15}
		W_3 (\psi) = 4 \kappa B D \sqrt{2} \left(  \frac { Dk(k-2)}{6 m}\right)^{\frac{k}{2-k}} \left( \frac{2-k}{2+k}\right) \left( \frac{\psi}{\sqrt{2m}} \right)^{\frac{2+k}{2-k}}+\frac{2\epsilon}{c^2} \left( \frac{\sqrt{2m}}{\psi} \right)^{2m},
	\end{equation}
	here $A,B,D,k = const $. 
Substitute the dependencies \( \psi(\chi) \) \eqref{1.1} and \( \phi(\chi) \)  \eqref{1.2} into the potentials of the chiral fields \( W_1(\phi) \) \eqref{13}, \( W_2(\phi) \) \eqref{14}, and \( W_3(\psi) \) \eqref{15}:

	\begin{equation}\label{z}
		W_1(\chi)=\frac {m(3m-1)}{Q^2} \chi^{k-2},
	\end{equation}
	
	\begin{equation}\label{x}
		W_2 (\chi)= 4 \kappa \sqrt {2m} D A \left( \frac { Dk(k-2)}{6 m}\right)^{\frac{k}{2-k}} \left( \frac{2-k}{2k}\right) Q^{\frac{2k}{k-2}} (\chi)^k,
	\end{equation}
	\begin{equation}\label{c}
		W_3 (\chi) = 4 \kappa B D \sqrt{2} Q \left( \frac{2-k}{2+k}\right) \chi^{\frac{2+k}{2}} +\frac{2\epsilon}{c^2 Q^{2m}} \chi^{m(k-2)}.
	\end{equation}

The total potential \( V_G(\chi) \) is determined as the sum of all potentials \eqref{z} - \eqref{c} and the potential of the matter field $U(\chi)=D\chi^k$:
	\begin{equation} \label{potSp}
		V_G(\chi)=A_1  \chi^{k-2} + A_2\chi^{\frac{2+k}{2}} +A_3 \chi^k +A_4 \chi^{m(k-2)},
	\end{equation}
where the constants \( A_i \) are:
\[
A_1 = \frac{m(3m-1)}{Q^2}, \quad A_2 = 4\sqrt{2} BDQ \left( \frac{2-k}{k+2} \right), \quad A_3 =  4\sqrt{2m} DA \left( \frac{2-k}{2k} \right), \quad A_4 = \frac{2 \epsilon}{c^2Q^{2m}}.
\]
To incorporate derivatives of the chiral fields into the action (4), we calculate their derivatives with respect to coordinates:
\begin{equation}\label{1.6}
		\phi_{,\mu}= \sqrt{2m} \left( 1 - \frac{k}{2}\right) \chi^{-1}\chi_{,\mu},
	\end{equation}
	\begin{equation}\label{1.7}
		\psi_{,\mu} =\sqrt{2}Q \left( 1- \frac{k}{2}\right) \chi^{-\frac {k}{2}}\chi_{,\mu}.
	\end{equation}
Substituting \eqref{1.3}, \eqref{potSp} - \eqref{1.7} into the action \eqref{1}, we obtain:

	\begin{equation}\label{deiSp}
		S=\frac{1}{\kappa}  \int d^4 x \sqrt{-g}\left[ \dfrac{R}{2}-\frac{1}{2}g^{\mu\nu} \omega (\chi) \chi_{,\mu} \chi_{,\nu} -V_G(\chi)\right],
	\end{equation}
	where the kinetic function $\omega (\chi) $ is: :
	\begin{equation} \label{omega}
		\omega (\chi)=  h_{11} \left( \dfrac{d\phi}{d \chi}\right) ^2 +h_{22}(\chi) \left( \dfrac{d\psi}{d \chi}\right) ^2 + 1.
	\end{equation}
	
The action \eqref{deiSp} is a special case of the action for generalized scalar-tensor gravity, discussed in \cite{25}. 
It should be noted that after transitioning to the model \eqref{deiSp}, \eqref{omega} with a single scalar field, the parameters \( m \) and \( c \) can be considered independent of power-law inflation and can take arbitrary values. 

Using the general form of the equations for generalized scalar-tensor gravity provided in \cite{24}, the system of equations for the model \eqref{deiSp} is obtained as:

	\begin{equation}\label{4}
		3H^2= \dfrac{ \omega(\chi)}{2} \dot{\chi}^2  +V(\chi),
	\end{equation}
	
	\begin{equation}\label{5}
		3H^2+2\dot{H} +\dfrac{ \omega(\chi)}{2} \dot{\chi}^2 - V(\chi) =0.
	\end{equation}
	
To find the solutions of the system \eqref{4} -\eqref{5}, we will use the Ivanov-Salopek-Bond method, which is described in detail in \cite{9,10}. 
Following this method, we assume the dependence of the Hubble parameter \( H \) on the field \( \chi \), and by using the sum of equations \eqref{4} -\eqref{5}, we obtain:

	\begin{equation}\label{8}
		 \frac{ d H (\chi)}{d \chi}=-\dfrac{ \omega(\chi)}{2} \dot{\chi},
	\end{equation}
	\begin{equation}\label{poten_F}
		V_G(\chi)	= - \dfrac{2}{3\omega(\chi)}  \left( \frac{ d F}{d \chi}\right)^2+F_*^2,
	\end{equation}
where \( F(\chi) \) is the generating function associated with the Hubble parameter:
\begin{equation}\label{Hub_GenF}
		H(\chi)=\sqrt{\dfrac{1}{3}} \left( F+F_* \right).
	\end{equation}

Thus, by specifying the generating function \( F(\chi) \), we find the Hubble parameter from \eqref{Hub_GenF} and the potential using formula \eqref{poten_F}. 
Next, the dependence of the field \( \chi \) on time is determined by solving the ordinary differential equation \eqref{8} for the given kinetic function \( \omega(\chi) \). 
Finally, the dependence of the Hubble parameter (and the scale factor) on time is established.

\subsection{Choice of the function \( \omega(\chi) \).} 

The kinetic function \( \omega(\chi) \) from \eqref{omega} will be considered for two particular cases, suitable for any value \( k \neq 2 \):

1. For \( \epsilon \neq 0 \), \( m = \frac{k}{k - 2} \), \( D=\frac{6c^{k-2}}{(k-2)^2} \left( -2\epsilon \left(1-\frac{k}{2}\right)^2 \right)^{\frac{2-k}{2}} \), the function \( \omega(\chi) \) takes the form:
\begin{equation} \label{omega_1}
		\omega (\chi)=-\frac{k(2-k)}{2 \chi^2}.  
	\end{equation}

2. For \( \epsilon  = -1 \), \( m = 1 \), \( c = 1 \), the function  \( \omega(\chi) \) takes the form:
\begin{equation} \label{omega_2}
		\omega (\chi)=1.
	\end{equation}

With this definition, the parameter \( D \) remains free.  
In this case, the potential \( V_G(\chi) \) from \eqref{potSp} takes the form:
\begin{equation}\label{pot_omega_2}
		V_G(\chi)=A_2\chi^{\frac{2+k}{2}}+A_3\chi^k.
	\end{equation}

 \section{Classes of Solutions}
 
	\subsection{Solutions for $V_G(\chi) = A_1  \chi^2 + A_2 \chi^3 + (A_3+A_4) \chi^4$ and $\omega (\chi)=\dfrac{4}{\chi^2}$}

The potential \( V_G(\chi) \) from \eqref{potSp}, under the conditions \( k = 4 \), \( m = 2 \), \( \epsilon= -1 \), and \( D = -3c^2/4\epsilon \), transforms into:
	\begin{equation}\label{poten}
		V_G(\chi) = A_1  \chi^2 + A_2 \chi^3 + (A_3+A_4) \chi^4
	\end{equation}
where the constants \( A_i \) simplify as follows: $ A_1 = 5c^2, \quad A_2 = -2Bc, \quad A_3 = \frac{3c^2(1 - 4A)}{4}, \quad A_4 = -\frac{c^2}{2}.$

The form of the function \( \omega(\chi) \) from \eqref{omega_1}, for \( k = 4 \), also simplifies:
\begin{equation}\label{37}
		\omega (\chi)=\frac{4}{\chi^{2}}.
	\end{equation}

To obtain the potential in the form of \eqref{poten}, the generating function is chosen as follows \cite{9}:
	\begin{equation}\label{F_a}
		F(\chi)= \sum_{k=0}^{p} \lambda_n \chi^n+F_*.
	\end{equation}

Substituting the generating function of the form \eqref{F_a} into the formula for the potential \eqref{poten_F}, and considering \( F_* = 0 \), \( p = 2 \), \( \lambda_0 = 0 \), $n=0,1,2$, we obtain:
	\begin{equation} \label{GPot}
		V_G(\chi)=\left(\dfrac{5\lambda_{1}^2}{6} \right) \chi^2 + \left(\dfrac{4\lambda_{1}\lambda_{2}}{3} \right) \chi^3+ \left(\dfrac{\lambda_{2}^2}{3} \right) \chi^4.
	\end{equation}

For \( \epsilon = -1 \), \( \lambda_1 = \pm c\sqrt{6} \), \( \lambda_2 = \pm \frac{3B}{2\sqrt{6}} \), and \( B^2 = 6c^2 (1-12A) \), the potential \eqref{GPot} matches the form of \eqref{poten}. Consequently, the generating function \eqref{F_a} for this case is:
\begin{equation}
		F(\chi)=\pm c\sqrt{6} \chi\pm \frac{3B}{2\sqrt{6}} \chi^2.
	\end{equation}

The Hubble parameter \( H(\chi) \) from \eqref{Hub_GenF}  takes the form:
	\begin{equation} \label{40}
H(\chi) = \pm c\sqrt{2} \chi \pm \frac{B}{2\sqrt{2}} \chi^2. 
	\end{equation}

Substituting the derivative of the Hubble parameter \eqref{40} with respect to \( \chi \) and function $\omega(\chi)$ \eqref{37} into equation \eqref{8}, we find the dependence \( t(\chi) \):
	\begin{equation}
t-t_* = \pm \frac{\sqrt{2}}{c} \left[  \frac{B}{2c} \ln \left(1 - \frac{2c}{B\chi}\right)+\chi^{-1} \right].
\end{equation}

    \subsection{Solutions for $V_G(\chi) = A_1 \chi^2$ and $\omega (\chi)=\dfrac{4}{\chi^2}$}
 
Consider the case of a weak-field potential \eqref{potSp}, where the third and fourth powers can be neglected.  
In this case:
	\begin{equation} \label{41}
V_G(\chi) = A_1 \chi^2, 
	\end{equation}
where \( A_1 = 5c^2 \).

Similar to the previous case, for this model, the conditions $\epsilon=-1$, \( k = 4 \), \( m = 2 \), \( c = \frac{2}{3} \), \( D = \frac{3c^2}{4} \) hold, and the value of the function \( \omega(\chi) \) corresponds to \eqref{37}. For the potential \eqref{41}, the generating function takes the form:
	\begin{equation} \label{42}
F(\chi) = \lambda_3 \chi.
\end{equation}

Substituting the function \eqref{42} into \eqref{poten_F} and considering \eqref{37}, we obtain:
	\begin{equation} \label{43}
V_G(\chi) = \frac{5\lambda_3^2}{6} \chi^2. 
	\end{equation}

It follows that for \( \lambda_3 = 6c \), the potential \eqref{43} is equal to the potential \eqref{41}. The Hubble parameter \eqref{Hub_GenF} in this case is:
	\begin{equation} \label{44}
		H(\chi) = c\sqrt{2} \chi 
	\end{equation}

The field \( \chi(t) \) for this case is determined from \eqref{8}:
\begin{equation}
\chi(t) = \frac{2}{c\sqrt{2} t}.
	\end{equation}	

The dependence of the Hubble parameter \eqref{44} on time and the scale factor of the model are as follows:
\begin{equation}
H(t) = \frac{2}{t}, \quad a(t) = a_0 t^2.
\end{equation}	

This solution corresponds to a specific case of power-law inflation in Friedmann cosmology, as presented, for example, in \cite{11} (see Table 3.1):
\[
H(t) = \frac{B}{3t} -\frac{A}{3B}, \quad \text{for  B= 6,   A= 0 }.
\]

\subsection{Solutions for $V_G(\chi) =  (A_1  + A_2 )\chi^4 $ and $\omega (\chi)=\frac{12}{\chi^2}$}

Consider the model for \( k = 6 \), \( m = \frac{3}{2} \), \( A = \frac{5\sqrt{3}c^6}{4\epsilon8^4} \), and \( D = \frac{3c^4}{512\epsilon^2} \). For this model, the function \( \omega(\chi) \) from \eqref{omega_1} becomes:
	\begin{equation} \label{45}
\omega(\chi) = \frac{12}{\chi^2}. 
\end{equation}

The potential \( V_G(\chi) \) from \eqref{potSp}, considering \( k = 6 \), takes the form:
\begin{equation} \label{46}
V_G(\chi) = (A_1  + A_2 )\chi^4, 
\end{equation}
where \( A_1 = \frac{21c^2}{32 \epsilon} \) and \( A_2 = -\frac{3\sqrt{2}Bc^2}{32 \epsilon} \).

This potential \eqref{46} corresponds to an approximation of the Higgs potential \cite{7}. The generating function is defined as:
\begin{equation}\label{47}
F(\chi) = \lambda_4 \chi^4. 
\end{equation}
Substituting functions \eqref{47} and \( \omega(\chi) \) into \eqref{potSp}, we find the potential:
	\begin{equation} \label{48}
V_G(\chi) = \frac{7\lambda_4^2}{9} \chi^4.
\end{equation}

The potential \eqref{48} matches \eqref{46} for \( \lambda_4 = \frac{3c}{4}\sqrt{\frac{3}{14\epsilon} (7-3\sqrt{2}B)} \). Thus, the Hubble parameter \eqref{Hub_GenF}, considering the generating function \eqref{48}, is:
	\begin{equation} \label{49}
H(\chi) = \frac{3c}{4}\sqrt{\frac{(7-3\sqrt{2}B}{14 \epsilon}} \chi^2
\end{equation}

The field \( \chi(t) \), determined from \eqref{8}, for this model is:
\[
\chi(t) = \sqrt{\frac{2\sqrt{14} \epsilon}{c\sqrt{(7-3\sqrt{2}B)}}} \cdot\frac{1}{\sqrt{t}}.
\]

Consequently, the Hubble parameter \eqref{49} and the scale factor for the model are:
\[
H(t) = \frac{2}{3t}, \quad a(t) = a_0 t^{3/2}.
\]

This solution also corresponds to a specific case of Friedmann cosmology presented in \cite{11}, Table 3.1:
\[
H(t) = \frac{B}{3t} -\frac{A}{3B}, \quad \text{for } B = 9/2, A = 0.
\]

\subsection{Solutions for $V_G(\chi) =  A_2 \chi^2 + A_3 \chi^6 $ and $\omega (\chi)=1$}

 Consider the case for \( k = 6 \), \( m = 1 \), and \( \omega(\chi) = 1 \). Then, the potential \eqref{potSp} takes the form:
\begin{equation} \label{50}
V_G(\chi) = A_2 \chi^2 + A_3 \chi^6, \quad (50)
\end{equation}
where \( A_2 = -B\sqrt{2D} \) and \( A_3 = \frac{D}{3} (3-4A\sqrt{2}) \).

For the potential \eqref{50}, the generating function is:
\begin{equation} \label{51}
F(\chi) = \lambda_5 \chi^5. 
\end{equation}

Substituting function \eqref{51} into equation \eqref{poten_F}, we obtain the potential:
\begin{equation} \label{52}
V_G(\chi) = -6\lambda_5^2 \chi^4 + \lambda_5^2 \chi^6. 
\end{equation}

This potential \eqref{52} equals the given potential \eqref{50} for \( \lambda_5 = \sqrt{\frac{D}{3}  (3-4A\sqrt{2})} \) and \( B = \sqrt{2D} (3-4A\sqrt{2})\). 

Similarly, the Hubble parameter and field \( \chi(t) \) are:
\[
H(\chi) =\sqrt{\frac{D}{3}  (3-4A\sqrt{2})} \chi^3, \quad \chi(t) = \frac{\sqrt{3}}{6 \sqrt{D(3-4A\sqrt{2})}} t^{-1}.
\]

The Hubble parameter and scale factor for the model are:
\[
H(t) = \frac{1}{72D  (3-4A\sqrt{2})t^3}, \quad a(t) = a_0 \left( -\frac{1}{36D  (3-4A\sqrt{2}) t^2}\right).
\]

This solution also corresponds to a specific case of Friedmann cosmology presented in \cite{11}, Table 3.1:
\[
H(t) = C \left[ \frac{B+4}{6CB}t \right]^{-\frac{B}{B+4}}, \quad \text{for } B = -6, C= \sqrt[4]{-\frac{1}{9D(3-4A\sqrt{2})36^3}}, \text{here }  D<0.
\]

\newpage

\section{Conclusion}\label{Section7}
 
A method is proposed based on the use of a particular case of an analytical solution in a multi-field cosmological model to establish a functional relationship between fields, which allows transforming a three-field model into a single-field one.

Using standard methods for calculating cosmological parameters in a single-field model, it becomes possible to perform calculations for such parameters in the tensor-multi-scalar gravity theory, which describes the epoch of early inflation in the Universe. The aim of this work was not to verify the model solutions against observational data. However, all models derived from the three-field model contain solutions that, for specific choices of free parameters, lead to well-known solutions previously presented in \cite{11}.

The presence of free parameters in the model gives hope for successful alignment with observational data. Several free parameters were obtained during the solution process, along with constraints on them:\\
- Part 5.1 – three free parameters: $c$, $B$, and $A$. A relationship between the conformal constants $A$ and $B$ was obtained here.\\
- Part 5.2 – one free parameter: $c$.\\
- Part 5.3– two free parameters: $c$, $B$.\\
- Part 5.4  – three free parameters: $A$, $B$, $D<0$, and a relationship between the conformal constants A and B was derived.

In case of discrepancies with observational data, quantum corrections at the early Universe stage, reflected in the Einstein-Gauss-Bonnet (EGB) gravity theory, can be considered. The transition to EGB gravity can be carried out without specific calculations, based on the method of I.V. Fomin presented in \cite{26}.

\section*{Acknowledgements}

The article was written within the framework of Additional Agreement No. 073-03-2024- 060/1 dated February 13, 
2024 to the Agreement on the provision of subsidies from the federal budget for financial support for the implementation 
of the state task for the provision of public services (performance of work) No. 073-03-2024-060 dated January 18, 2024, 
concluded between the Federal State Budgetary Educational Institution of Higher Education “UlSPU I.N. Ulyanov” 
and the Ministry of Education of the Russian Federation.

\end{document}